# Stock mechanics: theory of conservation of total energy and predictions of coming short-term fluctuations of Dow Jones Industrials Average (DJIA)

Çağlar Tuncay
Department of Physics, Middle East Technical University
06531 Ankara, Turkey
caglart@metu.edu.tr

**Abstract**

Predicting absolute magnitude of fluctuations of price, even if their sign remains unknown, is important for risk analysis and for option prices. In the present work, we display our predictions about absolute magnitude of daily fluctuations of the Dow Jones Industrials Average (DJIA), utilizing the original theory of conservation of total energy, for the coming 500 days.

**Introduction**

One may define total energy ($E_T$) for prices $\chi(t)$ (taking mass equal to unity) as

$$E_T = \tfrac{1}{2} v^2 + U , \qquad (1)$$

where v is the usual speed, i.e., $v(t)=d\chi(t)/dt$, and U is the potential energy. In literature a few number of references, which utilize the idea of potential energy is found [1-10]. J-P. Bouchaud, and R. Cont [1] developed a quadratic form within a Langevin approach, and in [2], J-P. Bouchaud studied a more complex expression involving a term with power three besides a quadratic one. In [3, 4] a potential energy corresponding to some states of random motion is phrased. An angle dependent potential is utilized to describe equilibrium and off equilibrium price states in [5]. K. Ide and D. Sornette studied oscillatory finite-time singularities in finance [6, 7], where the time derivative of speed was written in terms of a linear combination of $\chi^n$ and $v^m$, with arbitrary powers (See Eqs. (11)-(13), (32), (33), in [6], and Eqs. (1)-(3) in [7].) There, no assumption about the conservation of the sum of kinetic and potential energies was made. They have obtained oscillations and up (down) trends by neglecting the speed and restoring terms, respectively. Recently, we described some aspects of potential energy for price oscillations in [8], in terms of buying and selling processes within a herding mechanism. More recently, we proposed and utilized conservation of sum of potential and kinetic energy to make some predictions on several world indices [9, 10], where a linear potential energy ($U=g\chi$) was also proposed for the current epoch of DJIA, S&P500, and NASDAQ. The corresponding equations of motion (obtained by taking the time time derivative of Eq. (1) and equating it to zero as $d^2\chi(t)/dt^2 = -g$, and integrating afterwards)

$$\chi(t) = \chi_0 + v_0 t - \tfrac{1}{2} g t^2 \qquad (2)$$

survived as valid for the behavior of the DJIA, S&P500, and NASDAQ[8,10], within the last nine months after we proposed it, as shown for DJIA in Fig. 1. Unit for price (value) is taken as local currency unit (lcu), and for time as (day).

In the present work DJIA will be focused on, because its time domain is the longest and volume is largest of all the world indices. One may see Fig. 1. and its caption for the explicit form of Eq. (2) and the parameters therein, where (and in other related figures and computations) real data in [11] is utilized up to 30 Dec 2005.

**Conservation of total energy and predicting daily fluctuations for future**

As shown empirically in [10], the general trends averaged over months or years gave a roughly conserved total energy. The corresponding equation of motion for the current epoch, Eq. (2), may be decorated for future's probable behavior of the price as

$$\chi^{future}(t) = \chi(t)\,(1 - \alpha(0.5-\delta(t))) \quad , \tag{4}$$

where $\chi(t)$ is the expression in Eq (2), and $\delta(t)$ is the random function with $0 \leq \delta(t) < 1$. In Eq(4) $\alpha$ is any scaling factor, which controls the maximum value of deviation of the probable future value of the index from Eq. (2), where the shape remains independent of $\alpha$. In reality $\alpha$ may also depend on the regulations of the exchange, since in some countries arbitrary fluctuations are not allowed and limited by law. $\alpha$ maybe taken as 0.3 for DJIA for reasonable results with $\pm 15\%$ daily variations (probable) at maximum. It is clear that perfection can never be reached with any prediction, yet one may try better randomization processes for better approximations. Utilizing Eq. (9) of [8], we obtained the results displayed in Figs. 2. a., b., where the initial date (i.e. the first day) is taken as 02 Jan 1998. In the upper part of Fig. 2. b., the original (one of the probable) results are displayed with a shifted vertical axis, in order to seperate the upper curve from the lower one. In the lower curve of Fig. 2. b., which involves results of the real data, fluctuations after t=1251 (day) have become smaller than they were before. For the term $1 < t < 625$ (day) the same effect is also present but less visible, where some fluctuations have occured with big amplitudes at a few number of days.

**Summary and conclusion**

With a linear potential energy for the current epoch of DJIA, energy conservation theory yields in equations of motion for the index in terms of long-term moving averages, which may be utilized by some random decoration processes for future. Then the resulting decorated index value may be used for short-term (day to day) fluctuations and for their averages. As predicted within the present formalism, the index will fall to 8,000's and below within the coming 500 days, as a result the fluctions will increase accordingly, as shown on the right hand side bottom corner of Fig 2. a.

**Acknowledgement**

**Figure captions**

**Figure 1.** DJIA, about the Apr-Sep2000 climax is described by a rise and a fall of the price in a gravity g = –0.01212 (lcu/day$^2$). The initial (shooting) speed at the beginning of 1995 is $v_{01}$= 10.17 (lcu/day). The price falls down after the maximum height and inelastically bounces back with $v_{02}$= 8.43 (lcu/day) in the same gravity, and rises up to 11,000's in accordance with the expression $\chi - \chi_0 = (v_0)^2/2g$, which is obtained by taking time derivative of Eq. (1) and equating it to zero and performing a few algebraic opererations afterwards. A recession, back to 8,000's and below is expected within the coming 500 days.

**Figure 2.** Excel graphics for the probable future fluctuations of DJIA, which are obtained in terms of Eq. (9) of [8].

**Figure 2. a.** Probable decoration of the index for the past (from 02 Jan 1998 up to 30 Dec 2005, covering 2001 days) and for the coming 500 days. The inset below involves <v*v>, where the amplitude depends on α of Eq. (4). Thin line is the index, dashed line is the equation of motion for a linear -gravitational- potential energy of Eq. (3), and thick line is the decoration involving randomness of Eq. (4).

**Figure 2. b.** Probable fluctuations of returns due to decoration displayed in Fig. 2. a. The vertical axis is shifted to 1000 deliberately, on the aim of clear displaying. Below is the same quantity obtained utilizing real data.[11]

**FIGURES**

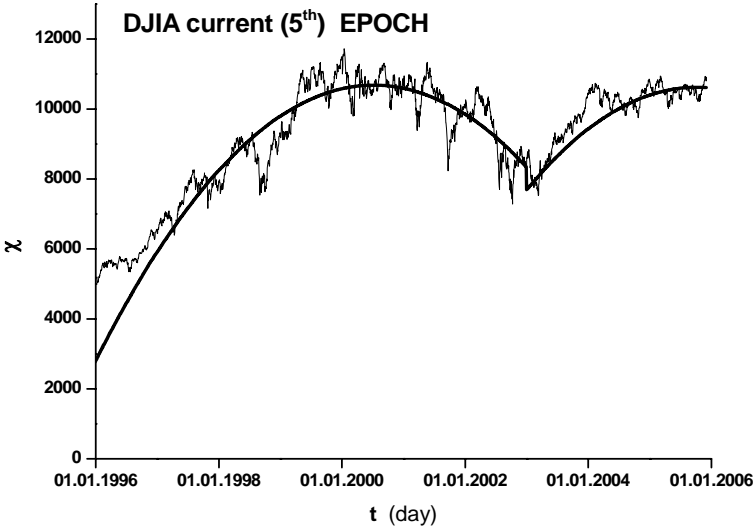

**Figure 1**

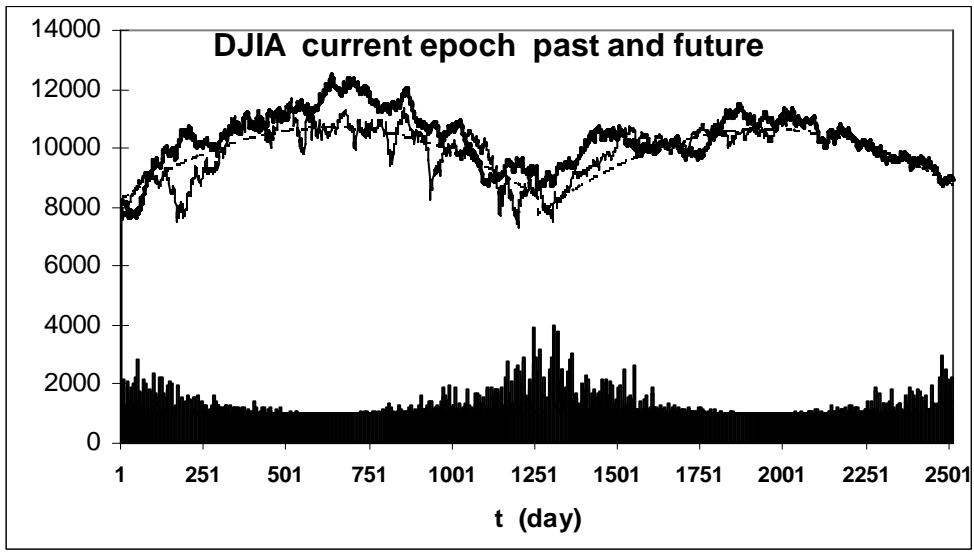

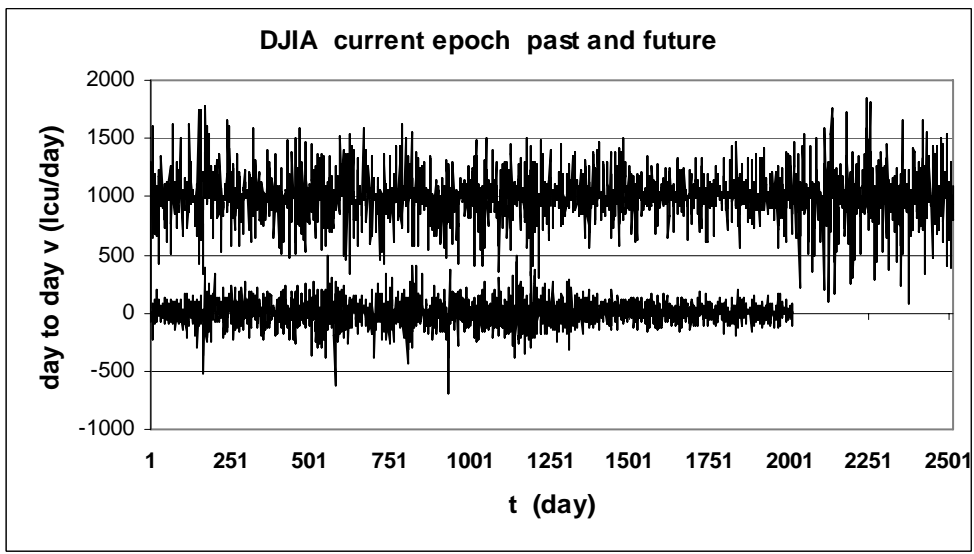

**Figure 2. a. and b.**